\documentclass{article}

\usepackage[preprint,nonatbib]{neurips_2023}

\usepackage[utf8]{inputenc} 
\usepackage[T1]{fontenc}    
\usepackage{hyperref}       
\usepackage{url}            
\usepackage{booktabs}       
\usepackage{amsfonts}       
\usepackage{nicefrac}       
\usepackage{microtype}      
\usepackage{xcolor}         
\usepackage{graphicx}
\usepackage{tabularx}
\usepackage{multirow}
\usepackage[linesnumbered,ruled,vlined]{algorithm2e}
\usepackage{makecell}
\usepackage{verbatim}
\usepackage{subfigure}
\usepackage{amsmath}
\usepackage{amssymb}
\usepackage{xspace}

\newcommand{\M}{\textsc{FedRKG}\xspace}

\title{\M: A Privacy-preserving Federated Recommendation Framework via Knowledge Graph Enhancement}

\author{%
  Dezhong Yao$^{1}$ \quad Tongtong Liu$^{1}$ \quad Qi Cao$^{2}$ \quad Hai Jin$^{1}$\\
  $^1$Huazhong University of Science and Technology \quad $^2$University of Glasgow \\
  \texttt{\{dyao,tliu,hjin\}@hust.edu.cn} \\
  \texttt{qi.cao@glasgow.ac.uk}
}

\begin{document}

\maketitle

\begin{abstract}
  \textit{Federated Learning} (FL) has emerged as a promising approach for preserving data privacy in recommendation systems by training models locally. Recently, \textit{Graph Neural Networks} (GNN) have gained popularity in recommendation tasks due to their ability to capture high-order interactions between users and items. However, privacy concerns prevent the global sharing of the entire user-item graph. To address this limitation, some methods create pseudo-interacted items or users in the graph to compensate for missing information for each client. Unfortunately, these methods introduce random noise and raise privacy concerns.
  In this paper, we propose \M, a novel federated recommendation system, where a global \textit{knowledge graph} (KG) is constructed and maintained on the server using publicly available item information, enabling higher-order user-item interactions. On the client side, a relation-aware GNN model leverages diverse KG relationships. 
To protect local interaction items and obscure gradients, we employ pseudo-labeling and \textit{Local Differential Privacy} (LDP).
Extensive experiments conducted on three real-world datasets demonstrate the competitive performance of our approach compared to centralized algorithms while ensuring privacy preservation. Moreover, \M achieves an average accuracy improvement of 4\% compared to existing federated learning baselines.
\end{abstract}

\section{Introduction}
\begin{figure}[t]
  \centering
  \includegraphics[width=\linewidth]{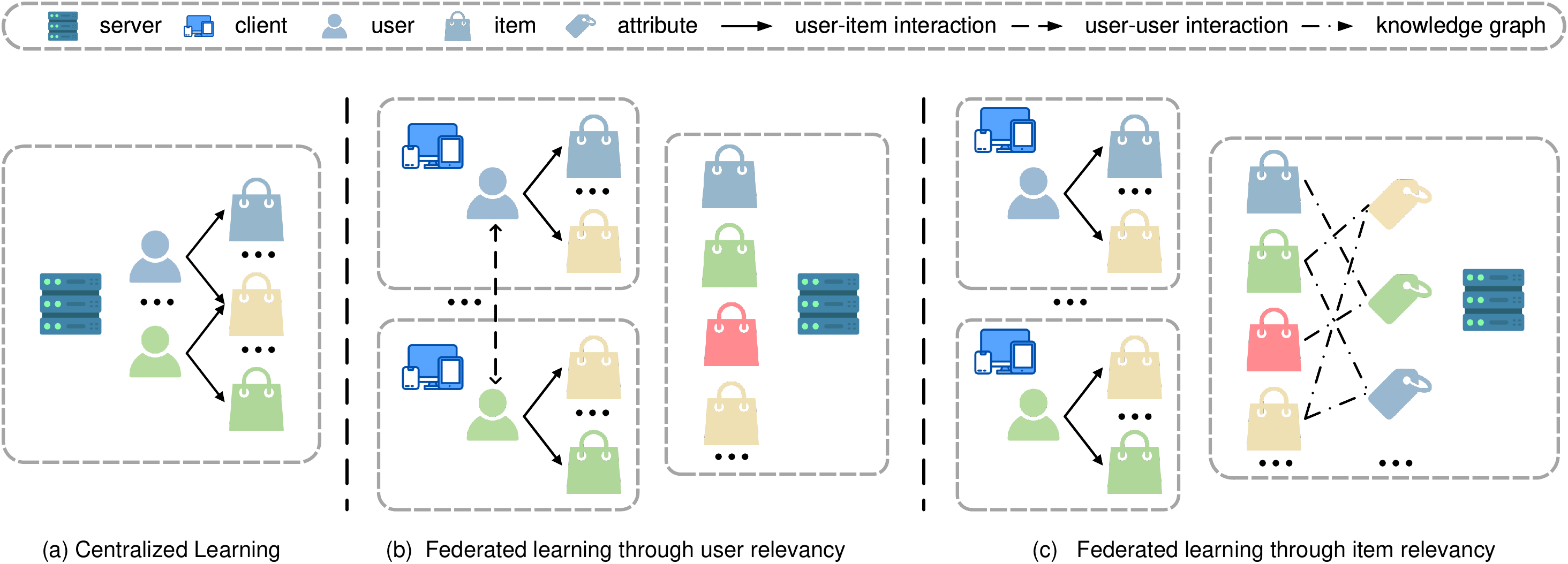}
  \vspace{-1.5em}
  \caption{Comparison of centralized learning, federated learning with enhanced user connections, and federated learning with enhanced project connections.}
  \label{fig:setting}
  \vspace{-1.5em}
\end{figure}

Recommendation systems are widely used in various domains, such as e-commerce and social recommendation, by alleviating users from the burden of sifting through vast amounts of data to discover suitable options~\cite{wang2019kgat}. 
These systems utilize user preferences and relevant information to provide personalized recommendations, making the process of finding relevant items more efficient and convenient~\cite{guo2020survey}.
However, the effectiveness of most recommendation methods heavily relies on centralized storage of user data~\cite{zhou2020graph}.
User data generated from software usage has the potential to enhance user experiences, deliver personalized services, and provide insights into user behavior~\cite{wu2022graph}.
Nevertheless, user data inherently includes user preferences and involves personal privacy. 
With the increasing awareness of privacy and the implementation of relevant regulations such as the \textit{General Data Protection Regulation} (GDPR)~\cite{voigt2017eu}, service providers may face growing challenges in centrally storing and processing user data, as shown in Fig.~\ref{fig:setting}(a).

The exclusive client access to local data leads to two challenges. 
Firstly, limited access to first-order interaction data hampers the effectiveness of the recommendation model.
Secondly, privacy-preserving mechanisms are required to ensure secure communication between the client and server. 
To address these challenges, FL is introduced into the recommendation system.
Existing works focus on the case of Fig.~\ref{fig:setting}(b), where recommendations are achieved by directly finding correlations between users. 
For example, FedMF~\cite{chai2020secure} and FedGNN~\cite{wu2021fedgnn} use only the local user-item interaction graph to find links between different users by \textit{collaborative filtering} (CF). 
However, incorporating various types of information in the conventional graph recommendation task can significantly improve the recommendation accuracy while changing the graph structure~\cite{wu2022graph}. 
Additionally, FeSoG~\cite{liu2022federated}  utilizes social networks as side information, adding direct connections between different users.
Nevertheless, this method requires the server to possess the complete social network, which is a type of private data that is difficult to obtain for most recommendation systems.
Furthermore, methods like FedGNN employ homomorphic encryption, which incurs substantial computational overhead and is not suitable as the primary encryption algorithm on edge devices with performance constraints.

To maximize the utilization of diverse data types while ensuring privacy protection on edge devices, we propose \M\footnote{The source code is available at: \url{https://github.com/ttliu98/FedRKG}}, a GNN-based federated learning recommendation framework. 
Unlike CF or direct construction of connections between users using privacy-sensitive information, \M leverages publicly available item information (e.g., appearance, attributes) to establish higher-order connections between different items, as shown in Fig.~\ref{fig:setting}(c).

The server firstly constructs and maintains \textit{knowledge graphs} (KGs) by utilizing publicly available item information.
Then, we employ on-demand sampling of KGs and distribute them to the client. Subsequently, we design a novel method to expand the local graph by merging KG subgraphs with the local user-item interaction graph, enabling the construction of high-order user-item interactions through KGs. Additionally, our framework introduces a request-based distribution mechanism. By obfuscating interaction items into request items, the server can efficiently distribute only the necessary request embeddings, significantly reducing communication overhead compared to previous methods while effectively protecting the privacy of raw interaction items. Simultaneously, we employ \textit{local differential privacy} (LDP) to protect all uploaded gradients, further enhancing the privacy of the federated learning process. 
Our approach has been extensively evaluated on three real-world datasets, demonstrating its competitive performance compared to centralized algorithms while ensuring privacy preservation.
Moreover, \M\ outperforms existing federated learning baselines, achieving an average accuracy improvement of approximately 4\%.
The major contributions of this work are summarized as follows:

\begin{itemize}
  \item To the best of our knowledge, we are the first to introduce a knowledge graph to enhance the performance of the federated recommendation system while protecting privacy.
  \item We introduce an algorithm for user-item graph expansion using KG subgraphs to improve local training.
  \item We propose innovative privacy-preserving techniques for interaction items, while simultaneously reducing communication overhead through strategic distribution of embeddings.
\end{itemize}

\section{Related Work}
\subsection{Knowledge Graph Based Recommendation}
In recent years, significant research has focused on recommendation systems that utilize \textit{Graph Neural Networks} (GNNs). 
GNNs have gained attention and popularity in recommendation systems due to their ability to learn representations of graph-structured data, which is well-suited for the inherent graph structures in recommendation systems. 
Knowledge graphs, as a typical graph structure, are often leveraged as side information in recommendation systems. 
By incorporating knowledge graphs, high-order connections can be established through the relationships between items and their attributes. 
This integration enhances the accuracy of item representations and provides interpretability to the recommendation results.
One type of method is integrating user-item interactions into KG. 
Methods like KGAT~\cite{wang2019kgat}, CKAN~\cite{wang2020ckan}, and MKGAT~\cite{sun2020multi} treat users as entities within KG, and relationships between users and items are incorporated as part of KG's relationships, too. 
This integration enables the merged graph to be processed using a generic GNN model designed for knowledge graphs. 
Another idea is employed by KGCN~\cite{wang2019knowledge} and KGNN-LS~\cite{wang2019knowledge2}, directly connecting KG to the user-item graph without any transformation. 
These methods utilize relation-aware aggregation and consider the user's preference for relationships when generating recommendations.

\subsection{Federated Learning for Recommendation System}
Federated learning is extensively utilized in privacy-preserving scenarios, as it ensures that the original data remains on local devices while allowing multiple clients to train a model together~\cite{JinBYDGYS23}. 
Considering the information required for recommendations, which includes users' preferences for items, the introduction of federated learning can help us prevent privacy breaches.
FedSage~\cite{zhang2021subgraph} and FKGE~\cite{peng2021differentially} focus on cross-silo federation learning, they are not suitable for protecting the privacy of individual users on client devices. 
FCF~\cite{ammad2019federated} and FedMF~\cite{chai2020secure} decompose the scoring matrix, retain user embeddings locally, and aggregate item embeddings on the server. 
FedGNN~\cite{wu2021fedgnn} utilizes homomorphic encryption for CF and protects the original gradients using pseudo-labeling and LDP. 
However, the computational requirements for homomorphic encryption pose challenges, particularly on performance-constrained devices.
In contrast to methods that do not leverage any side information, FeSoG~\cite{liu2022federated} introduces social networks to establish connections between users. 
Unfortunately, in many recommendation scenarios such as e-commerce, service providers do not offer social services, and social network information is considered private. Therefore, the lack of user connection on the server in Fig.~\ref{fig:setting}(b), like a social network, restricts the method's ability to generalize~\cite{10109123}. 
Currently, there is a scarcity of federated learning algorithms that effectively utilize side information for cross-device scenarios.

\section{Federated Recommendation with Knowledge Graph Enhancement}
\subsection{Problem Definition}
User-item interactions can be represented by a typical bipartite graph $\mathcal{G}=(\mathcal{U}, \mathcal{T}, E)$, where $\mathcal{U}=\left\{u_1, u_2, \ldots, u_N\right\}$ and $\mathcal{T}=\left\{t_1, t_2, \ldots, t_M\right\}$ represent a set of users and items of size $N$ and $M$, respectively. 
To describe the set of edges $E$, an interaction matrix $\mathbf{Y} \in \mathbb{R}^{M \times N}$ is employed. In particular, $y_{ut}$ takes on the value 1 if an interaction exists in the user's history, and 0 otherwise. 

For federated recommendation, each client $c_i$ owned by corresponding user $u_i$ can only access the interaction graph $\mathcal{G}_i$ stored locally, containing a set of items $\mathcal{T}_i$ that have been interacted with. Each $\mathcal{G}_i$ is a subgraph of the global interaction graph $\mathcal{G}$.

In addition to the client-side data, the server maintains a knowledge graph $K$, which is represented as a series of triples $\{(h, r, t) \mid h, t \in \mathcal{E}, r \in \mathcal{R}\}$. 
The entities $h$ and $t$ each refer to the head and tail, respectively, within the specific combination denoted by each triple, both belonging to the set of entities $\mathcal{E}$. 
The relationship $r$ represents the connection between two distinct entities, belonging to a set of relations $\mathcal{R}$.

Our goal is to train a generalized GNN model using the local bipartite graphs $\mathcal{G}_i$ and the knowledge graph $K$ while preserving user privacy. The model predicts the probability $\hat{y}_{ut}$ that a user $u$ will be interested in an unexplored item $t$.

\begin{figure}[t]
  \centering
  \includegraphics[width=\linewidth]{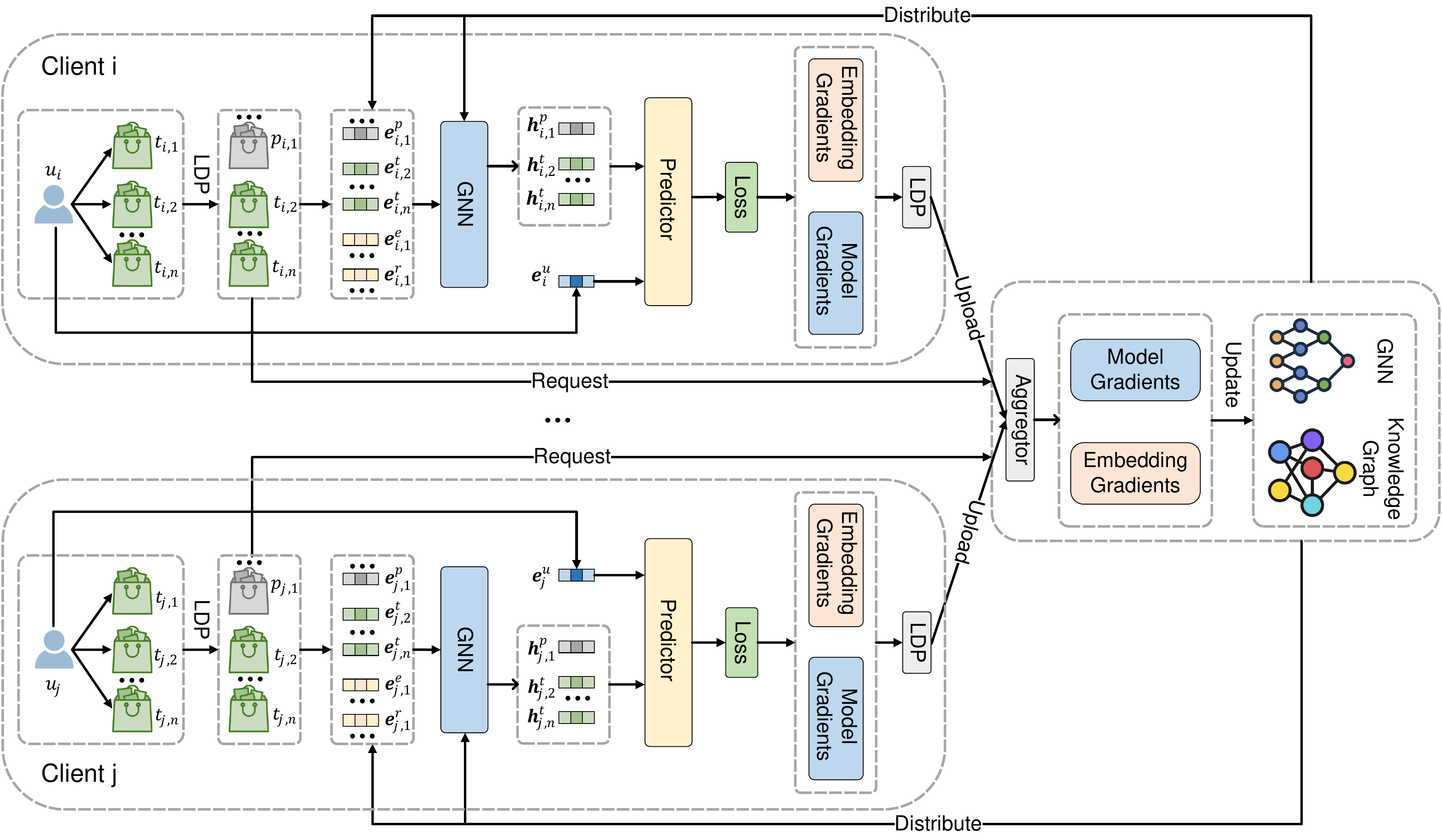}
  \vspace{-1.6em}
  \caption{The framework of \M.}
  \vspace{-1.5em}
  \label{fig:framework}
\end{figure}

\subsection{Framework Overview}
To enable privacy-preserving recommendation tasks across diverse private devices, we introduce a federated learning framework, in Fig.~\ref{fig:framework}, based on the knowledge graph named \M.
In the proposed framework, the client-server architecture is adopted. 
The client, which is the user's private device, is responsible for training a local graph neural network model. 
The server, on the other hand, is responsible for aggregating the models and embeddings, maintaining the knowledge graph, and constructing higher-order connections between clients.

\begin{algorithm}[t]
\label{alg:Fed}
\SetAlgoLined
\KwIn{
Neighbor sampling size $K$; embedding size $d$; depth of receptive field $H$; 
learning rate $\eta$; 
client number $N$; item number $M$; 
pseudo items $p$; $(0,1)$ flipping $q$;LDP parameter $\delta$, $\lambda$;
knowledge graph $K$; 
clients local graph $\left\{\left.\mathcal{G}_n\right|_{n=1} ^N\right\}$ }
\KwOut{GNN parameters and KG embeddings $\theta$,user embeddings $\left\{\left.\mathbf{e}_{u}^*\right|_{n=1} ^N\right\}$  }
Initialize $\theta$, $K$,$\left\{\left.\mathbf{e}_{u}^*\right|_{n=1} ^N\right\}$;

\While{\M not converged}{
Randomly select a subset $\mathcal{N}$ from $N$ randomly;

// client

\For{each client $n \in \mathcal{N}$}{
$\mathcal{T}_n^{\prime} \gets$ GenerateRequestItems$(\mathcal{G}{n}, p, q)$;

$\theta, \mathcal{G}_n \gets$ Request($\mathcal{T}_n^{\prime}$)

$g_n \gets$ LocalTrain($\theta, \mathcal{G}_n$)

$\Tilde{g}_n \gets$LDP($g_n$)

Upload($\Tilde{g}_n$)
}

// server

\For{each client $n \in \mathcal{N}$}{

$\mathcal{T}_n^{\prime} \gets$ReceiveRequest()

$\mathcal{G}_n \gets$ GetSubKG($\mathcal{T}_n^{\prime}$)

Distribute($\theta,\mathcal{G}_n$)

$\Tilde{g_n} \gets$ReceiveGrad()
}

$\overline{g} \gets$Eq.(\ref{eq:grad_sum})

$\theta \gets$Eq.(\ref{eq:update})
}
\caption{\M}
\end{algorithm}

The entire workflow is summarized in Algorithm~\ref{alg:Fed}, which concisely represents the complete workflow.

\subsection{Client Design}
In our framework, the client plays a crucial role in two tasks. 
First, it is responsible for ensuring the confidentiality of the user's private information during the communication process with the server, which is achieved through privacy-preserving algorithms. 
Second, the client utilizes the embeddings and models provided by the server to expand the local user-item graph and train the local model. 

Based on the knowledge graph shared by the server, we design a novel method to expand the local subgraph.
During the request phase, the client applies a privacy protection mechanism to the interaction items $\mathcal{T}_n$, generating obfuscated request items $\mathcal{T}_n^{\prime}$.
These request items are then transmitted to the server. 
The client receives a GNN model and a knowledge subgraph that includes the request items and some of their neighboring entities in the complete KG. By merging this knowledge subgraph with local user-item interaction, the client generates a graph for local training.
This approach guarantees the privacy of the user's interaction records by never disclosing them to the server, while also allowing the client to obtain more item-related information for training, thus indirectly enabling the construction of higher-order connections through knowledge subgraph.

\begin{figure}[t]
  \centering
  \includegraphics[width=0.7\linewidth]{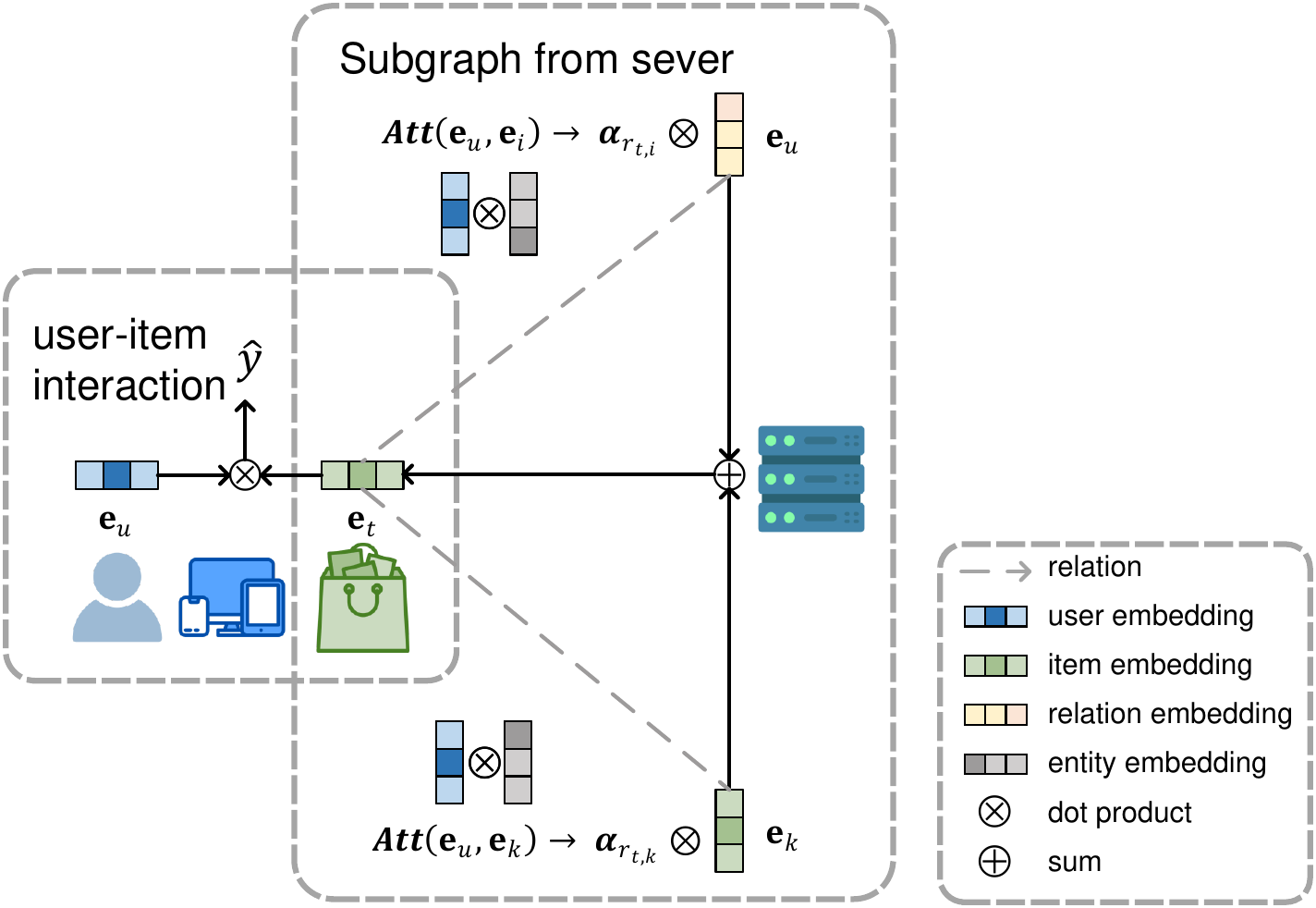}
  \vspace{-0.8em}
  \caption{Relation-aware aggregation in client.}
  \label{fig:agg}
  \vspace{-1.5em}
\end{figure}

Once the aggregated global model is received,  the client proceeds to update its local model and initiates a training process.
We use a relation-aware GNN as a recommendation model~\cite{cao2022geometric} that conforms to the message-passing paradigm~\cite{JinBYDGYS23}, as shown in Fig.~\ref{fig:agg}. For a given user $u$, entity $e_i$, $e_j$, and $r_{i, j}$ as the relation between two entities, we follow node-wise computation at step t+1:
\begin{equation}
    x_{i}^{(t+1)}=\phi\left(x_{i}^{(t)}, \rho\left(\left\{m_{r_{i,j}}^{(t+1)}:(u, e_j,r_{u,v}) \in \mathcal{E}\right\}\right)\right)
\end{equation}
where $x_i^t \in \mathbb{R}^d$ is embedding of entity $e_i$ in step $t$. We utilize a simple summation operation as the reduce function $\rho$ and directly replace the original embeddings with the aggregated results as the reduce function, denoted as $\phi$. $e_j$ sends a relationship-aware message $m_{r_{i,j}}$ to its neighbor:
\begin{equation}
    m_{r_{i,j}}=\alpha_{r_{i, j}}^{u}x_j
\end{equation}
where the attention score $\alpha_{r_{i, j}}^{u}$ between user $u$ and relation $r_{i,j}$ is derived from the following formula:
\begin{equation}
    s_{r_{t,i}}^{u}=score(\mathbf{e}_u, \mathbf{e}_{r_{t,i}})
\end{equation}
\begin{equation}
    Att(e_u,e_i)=\alpha_{r_{t, i}}^{u}=\frac{\exp \left(\ s_{r_{t,i}}^{u}\right)}{\sum_{i^{\prime} \in \mathcal{N}(t)} \exp \left(\ s_{r_{t,i^{\prime}}}^{u}\right)}
\end{equation}

We calculate an attention score using a score function (e.g. inner product) and then normalize it.
After obtaining the final embedding $x_t$ of item $t$, we calculate the prediction $\hat{y}$ by a readout function and then train this GNN model using BCE as the loss function. Finally,  client uploads encrypted gradient to server.

\subsection{Server Design}
Similar to clients, the server performs distinct tasks that are mainly distributed across two phases. 
Firstly, the server's primary responsibility is to respond to the client's requests. 
Based on the requested items, the server utilizes the knowledge graph to sample a subgraph that corresponds to a specific client. 
The subgraph comprises two key components, namely the structural information in the form of triples, and the feature information, represented by the embedding of entities and relations. 
Subsequently, the server shares the subgraph, together with the global model, with the client.
Secondly, the server needs to receive all gradients of local models and embeddings uploaded by clients.
These gradients are then aggregated and used to update the global model and knowledge graph.

In each communication round, the server activates $\mathcal{N}$ clients.
After receiving request items from those clients, server randomly samples a set of neighbors, denoted as $S(t) \triangleq \{e | e \sim N (t)\}$, for the request item $t$. Here, $|S(v)| = K$ represents the fixed size when sampling, and $N(t)$ represents immediate neighbors for item $t$. In our framework, $S(v)$ is also referred to as the (single-layer) receptive field of item $t$.
Repeat the above sampling several times to obtain $\mathcal{G}_i$ containing $n$ iterations and then distribute it to the client along with the parameters $\theta$, consisting of the model parameters $\theta^m$ and all embeddings of entities and relations in $\mathcal{G}_i$ denoted by $\theta^e$.
Finally, it receives the local gradients $\tilde{g}_i$ of these clients and aggregates them as follow:

\begin{equation}
\overline{\mathrm{g}}=\frac{\sum_{n \in \mathcal{N}}\left|\mathcal{T}_n^{\prime}\right| \cdot \tilde{\mathrm{g}}^{n}}{\sum_{n \in \mathcal{N}}\left|\mathcal{T}_n^{\prime}\right|}
\label{eq:grad_sum}
\end{equation}

After aggregation, the server updates all parameters $\boldsymbol{\theta}$  with gradient descent as:

\begin{equation}
\theta^{*}=\theta-\eta \cdot \bar{g}
\label{eq:update}
\end{equation}
where  $\eta$ is the learning rate. 

\subsection{Privacy-Preserving Communication}

\subsubsection{User privacy}
Within our proposed framework, user-related privacy pertains primarily to user embedding. Traditional embedding-based recommendation algorithms can derive both user and item embeddings and generate user-specific recommendations through a straightforward readout operation. However, user embeddings comprise the user's preference characteristics, which can lead to a compromise of their privacy.
In the federated learning scenario where the server does not have access to the raw data, to avoid exposing user preferences directly to the server, it is obvious that we need to keep the user embeddings on the client side and isolate them from the server. 
Clients can simply protect user-related privacy by refraining from uploading user embeddings after the training phase.

\subsubsection{Interaction privacy}
The interaction between users and items is considered highly sensitive information, susceptible to potential leaks during two stages. 
Firstly, due to the large size of the knowledge graph for items and limited transmission bandwidth, it is not practical to distribute all embeddings to client similar to FedGNN and FedSoG.
Instead, we aim to complete the entire training process through the limited distribution of embeddings. However, this presents a challenge in determining which embeddings should be distributed by the server. 
Server can not explicitly obtain the required embeddings, as this would mean that it has access to the client's real interaction item. 
Therefore, we need to obfuscate the original interaction items to obtain encrypted request items, which can then be sent to the server to sample the corresponding subgraph required for training.

We have designed a \textit{local differential privacy}(LDP) mechanism to generate request items from the interaction items. 
Specifically,  user-item interaction for user $u$ can be represented as a set $\{(t_i, y_{ui}) \mid y_{ui} \in\{0,1\},i=1,2, \ldots, n\}$. 
This collection contains $|\mathcal{T}|$ elements, each of which is a binary, the first of which is an item and the second is either $0$ or $1$, indicating whether the user interacted with the item. 
Let the query for the $t_i$  be $y_{ui}$, then the interaction can be privacy-preserving using an $\epsilon$-LDP algorithm. 
The privacy budget $\epsilon$ indicates the maximum acceptable loss of privacy.
Let the interaction for each item satisfy $\epsilon$-LDP, and we have:
for any item, keep the original interaction value with probability $\frac{e^\epsilon} {e^\epsilon+1}$ and invert it to another value with $\frac{1}{e^\epsilon+1}$ (0,1 flipping).

A potential privacy concern with the widely used pseudo-labeling method in previous work is that the interacted item will always generate gradients, even if pseudo-labeling is used. 
Additionally, the pseudo-labeling method applied during the training phase does not effectively reduce the communication overhead associated with distributing embeddings. 
To address this issue, we first sample several non-interactive samples, mix them with real interaction items, and further obfuscate them by the above LDP method to achieve privacy protection.

\subsubsection{Gradients privacy}
Ensuring the privacy of users' sensitive information is a critical concern when maintaining a knowledge graph and updating the global model in a federated recommendation system. 
In each communication round, the server needs to aggregate gradients of entity embeddings, relational embeddings, and GNN models from different clients. However, it has been demonstrated, as exemplified by FedMF, that uploading users' gradients in consecutive steps can lead to the inadvertent exposure of sensitive data, such as users' ratings.
Therefore, we need to obfuscate gradients to protect user privacy.
However clients of recommendation systems, such as mobile devices, often have limited computational capabilities~\cite{tian2022fedbert}, and computationally intensive methods like homomorphic encryption may not be practical to implement on such devices. 
To tackle this, we employ LDP by injecting random noise into the local gradients before uploading them to the server. 
This approach effectively protects row gradients without compromising the accuracy of the model. 
Moreover, it helps ensure that the computational overhead remains manageable and within acceptable limits.

To be more specific, gives all gradients as $g_n=\{g_{n}^{e}, g_{n}^{m}\}=\frac{\partial \mathcal{L}_{n}}{\partial \theta}$, where $\mathcal{L}_{n}$  denotes loss of client $n$, the LDP is formulated as:

\begin{equation}
\tilde{\mathbf{g}}_{n}=\operatorname{clip}\left(\mathbf{g}_{n}, \delta\right)+\text { Laplacian }(0, \lambda)
\end{equation}
where $\tilde{\mathbf{g}}_{n}$ is the encrypted gradient, $\operatorname{clip}(x, \delta)$ denotes the gradient clipping operation with a threshold $\delta$ to limit $x$ and prevent the gradient from being too large, after which we add to the gradient a mean value of 0 and an intensity of $\lambda$ of Laplacian noise, denoted by Laplacian $(0, \lambda)$. This results in a $\epsilon$-LDP, where the privacy budget $\epsilon$ is $\frac{2\delta}{\lambda}$.

\section{Experiment}

\subsection{Datasets}
In order to ensure the robustness of the algorithm, we aim to test the overall performance of the framework on a variety of datasets with different sizes, sparsity, and domains. Therefore, we have selected the following real-world datasets:

\begin{itemize}
  \item \textbf{MovieLens-20M}~\cite{harper2015movielens} contains five-star ratings from MovieLens, a movie recommendation service, as of 2019. Each user in the dataset has provided a minimum of 20 ratings (ranging from 1 to 5) on the MovieLens website.
  \item \textbf{Book-Crossing}~\cite{ziegler2005improving} contains user ratings (ranging from 0 to 10) of books extracted from the Book-Crossing community in 2004. In this dataset, a rating of 0 indicates an implicit interaction between the user and the book.
  \item \textbf{Last.FM}~\cite{cantador2011second} contains musician listening recodes from the Last.FM music streaming service. We consider artists as items and the number of listens as ratings. In particular, we utilize the HetRec 2011 version in our study.
\end{itemize}

To adapt the dataset for the recommendation task in a federated learning environment, several steps are taken. 
Firstly, only the user-item interactions are retained from the original dataset, while other data are discarded. 
Then, the publicly available Microsoft Satori is utilized to create a knowledge graph by selecting triples with a confidence level greater than 0.9, where the tail corresponds to items in the dataset. 
Interactions, where the item is not present in the knowledge graph, are subsequently removed.
Next, these three datasets are transformed into implicit feedback. 
We consider all artists listened to in Last.FM, all books with ratings present in book-cross, and all movies with ratings greater than or equal to 4 stars in MovieLens, as positive feedback.
Conversely, items not meeting these criteria are treated as negative feedback.
Lastly,  since the original recommendation dataset already contains user information, each user's data is assigned to the corresponding client to generate a federated learning dataset. 
Details of the dataset are shown in Table~\ref{tab:hyperparameters}

\begin{table}[t]
  \centering
\caption{Dataset basic information and hyperparameters, notation is consistent with Algorithm~\ref{alg:Fed}. }
  \label{tab:hyperparameters}
    \begin{tabular}{c|ccc}
    \hline
    & MovieLens-20M & Book-Crossing & Last.FM \\
    \hline
    users & 138,159 & 19,676 & 1,872 \\
    items & 16,954 & 20,003 & 3,846 \\
    interactions & 13,501,622 & 172,576 & 42,346 \\
    entities & 102,569 & 25,787 & 9,366 \\
    relations & 32 & 18 & 60 \\
    KG triples & 499,474 & 60,787 & 15,518 \\
    \hline
    $K$ & 4 & 8 & 8 \\
    $d$ & 32 & 64 & 16 \\
    $H$ & 2 & 1 & 1 \\
    \hline
    $\lambda$ & $10^{-7}$ & $2 \times 10^{-5}$ & $10^{-4}$ \\
    $\eta$& $2 \times 10^{-2}$ & $2 \times 10^{-4}$ & $5 \times 10^{-4}$ \\
    $\mathcal{N}$& 32768 & 64 & 32 \\
    \hline
    \end{tabular}
\end{table}

\subsection{Baselines}
We compare the proposed \M\  with the following baselines, in which the first two baselines are KG-free while the rest are all KG-aware methods. Hyper-parameter settings for baselines are introduced in the next subsection.
\begin{itemize}
    \item \textbf{SVD}~\cite{koren2008factorization} is a classical CF recommendation algorithm based on matrix decomposition. Here we use an unbiased version.
    \item \textbf{LibFM}~\cite{rendle2012factorization} is a method based on Factorization Machines that captures the similarity between features
    \item \textbf{PER}~\cite{yu2014personalized}  is an algorithm based on a personalized attention mechanism and constructs a Meta-path between users and items through a heterogeneous graph (KG).
    \item \textbf{CKE}~\cite{zhang2016collaborative} is a knowledge graph-based collaborative embedding recommendation algorithm that combines data from CF and other modalities.
    \item \textbf{RippleNet}~\cite{wang2018ripplenet} is a memory-network-like approach that simulates and exploits the ripple effect between users and items to propagate information on the knowledge graph
    \item \textbf{KGCN}~\cite{wang2019knowledge} is a KG-based method, that achieves efficient recommendations by merging KG and CF data.
    \item \textbf{FedMF}~\cite{chai2020secure} is a recommendation algorithm based on matrix decomposition while protecting privacy through an encryption mechanism.
    \item \textbf{FedGNN}~\cite{wu2021fedgnn} is a GNN-based recommendation algorithm that uses homomorphic encryption for aggregation and protects the original gradient by differential privacy and pseudo-labeling.
\end{itemize}

\subsection{Experimental Settings}
Table~\ref{tab:crt} shows the hyperparameter for the experiments. 
We split the datasets into training, validation, and testing sets in a 6:2:2 ratio. AUC and F1 scores are used as evaluation metrics for \textit{click-through rate} (CTR) prediction.

For the Last.FM, Book-Crossing, and MovieLens-20M datasets, the SVD method is applied with imensions $(8, 8, 8)$ and learning rates $(0.1, 0.5, 0.5)$. For LibFM, the dimensions are $(8, 1, 1)$. PER utilizes the user-item-attribute-item meta-path, with dimensions $(64, 128, 64)$ and learning rates $(0.1, 0.5, 0.5)$. The learning rates for KG in CKE are $(0.1, 0.1, 0.1)$, while the dimensions are $(16, 4, 8)$ and the H values are $(3, 3, 2)$. RippleNet's dimensions are $(16, 4, 8)$, H values are $(3, 3, 2)$, learning rates are $(0.005, 0.001, 0.01)$,  regularization parameters $\lambda_1$ are $(10^{-5}, 10^{-5}, 10^{-6})$, and  $\lambda_2$  are $(0.02, 0.01, 0.01)$. Other hyperparameters remain the same as in the original papers, and the federated learning settings are consistent with this paper.

\begin{table}[t]
  \centering
  \caption{Results for CRT prediction. KGCN achieves the best AUC among the first five centralized learning methods. Our method performs best in the next three federal learning methods, while the gap with KGCN is acceptable.}
  \label{tab:crt}
  \resizebox{1.0\linewidth}{!}{
    \begin{tabular}{c|cccccc}
        \hline
        \multirow{2}{*}{Model} & \multicolumn{2}{c}{MovieLens-20M} & \multicolumn{2}{c}{Book-Crossing} & \multicolumn{2}{c}{Last.FM} \\
        & \multicolumn{1}{c}{AUC} & \multicolumn{1}{c}{F1} & \multicolumn{1}{c}{AUC} & \multicolumn{1}{c}{F1} & \multicolumn{1}{c}{AUC} & \multicolumn{1}{c}{F1} \\
        \hline
        SVD & 0.952($\pm0.013$) & 0.909($\pm0.014$) & 0.665($\pm0.058$) & 0.628($\pm0.051$) & 0.760($\pm0.026$) & 0.688($\pm0.022$) \\ 
        LibFM & 0.960($\pm0.018$) & 0.907($\pm0.024$) & 0.692($\pm0.046$) & 0.619($\pm0.063$) & 0.779($\pm0.019$) & 0.711($\pm0.011$) \\ 
        PER & 0.824($\pm0.119$) & 0.780($\pm0.121$) & 0.611($\pm0.101$) & 0.557($\pm0.100$) & 0.627($\pm0.125$) & 0.593($\pm0.107$) \\ 
        CKE & 0.918($\pm0.050$) & 0.866($\pm0.056$) & 0.673($\pm0.057$) & 0.607($\pm0.055$) & 0.739($\pm0.044$) & 0.669($\pm0.046$) \\ 
        RippleNet & 0.964($\pm0.010$) & 0.909($\pm0.020$) & 0.712($\pm0.023$) & 0.648($\pm0.032$) & 0.777($\pm0.016$) & 0.699($\pm0.015$) \\ 
        KGCN & \underline{0.978}($\pm0.002$) & \underline{0.932}($\pm0.001$) & \underline{0.738}($\pm0.003$) & \underline{0.688}($\pm0.006$) & \underline{0.794}($\pm0.002$) & \underline{0.719}($\pm0.003$) \\
        \hline
        FedMF & 0.865($\pm0.012$) & 0.852($\pm0.015$) & 0.657($\pm0.039$) & 0.605($\pm0.060$) & 0.720($\pm0.018$) & 0.660($\pm0.013$) \\ 
        FedGNN & 0.939($\pm0.011$) & 0.891($\pm0.021$) & 0.671($\pm0.024$) & 0.620($\pm0.037$) & 0.753($\pm0.014$) & 0.681($\pm0.028$) \\ 
        \M & \textbf{0.970($\pm0.002$)} & \textbf{0.919($\pm0.002$)} & \textbf{0.724($\pm0.004$)} & \textbf{0.667($\pm0.006$)} & \textbf{0.785($\pm0.004$)} & \textbf{0.708($\pm0.002$)} \\ 
        \hline
    \end{tabular}
    }
\end{table}

\subsection{Overall Comparison}
\begin{figure}[ht]
  \centering
  \subfigure[MovieLens-20M]{
    \includegraphics[width=0.31\textwidth]{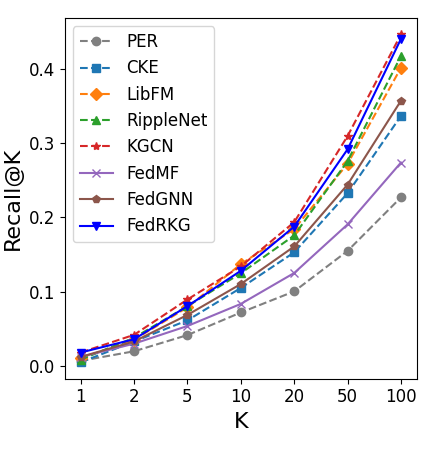}
    \label{fig:subfig1}
  }
  \subfigure[Book-Crossing]{
    \includegraphics[width=0.31\textwidth]{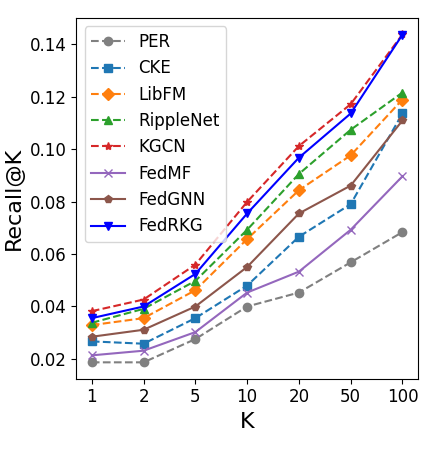}
    \label{fig:subfig2}
  }
  \subfigure[Last.FM]{
    \includegraphics[width=0.31\textwidth]{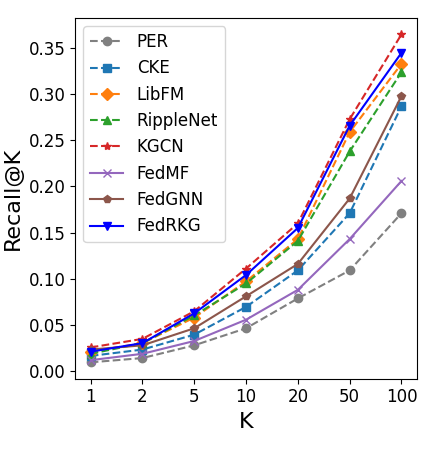}
    \label{fig:subfig2}
  }\\
  \vspace{-1.0em}
  \caption{Results for top-K recommendation. The dashed line represents centralized learning, while the solid line represents federated learning. Our method surpasses all federated baselines and, furthermore, achieves competitive results compared to centralized learning.}
  \label{fig:top-k}
  \vspace{-1.5em}
\end{figure}

We conduct a comprehensive comparison of multiple models under various settings. Given the dataset's specific characteristics, only including knowledge graphs and user-item graphs, many federated learning algorithms simplify to FedGNN in this dataset. Therefore, we select FedGNN and FedMF as the baseline methods, representing GNN and matrix decomposition approaches in federated learning. 
The experimental results for CTR prediction are presented in Table~\ref{tab:crt}, while Fig.~\ref{fig:top-k} illustrates the outcomes of top-k recommendation. Based on those results, we draw the following conclusions:
\begin{itemize}
    \item On the one hand, GNN-based algorithms, such as KGCN and \M, outperform matrix decomposition-based algorithms like SVD and FedMF. 
    This is due to the superior performance of GNNs in automatically capturing user preferences and enabling the spreading of user or item embeddings to neighboring nodes. 
    On the other hand, algorithms that require manual design such as meta-paths for PER and \textit{knowledge graph embedding} (KGE) method for CKE, often underperform due to the complexity of graph data.
    \item The experimental results consistently demonstrate that the appropriate utilization of additional side information can significantly improve the accuracy of recommendation systems. 
    For example, KGCN and RippleNet outperform other centralized algorithms regarding both AUC and F1 metrics, while \M, as a knowledge graph-based algorithm, performs best in federated learning. However, it should be noted that not all methods that leverage side information deliver satisfactory outcomes. This holds true for methods like PER and CKE, which encounter difficulties in effectively harnessing side information.
    \item Knowledge graphs are well-suited for integration into recommendation systems as side information, especially using GNNs, given their inherent graph structure and the ability to combine multi-domain knowledge. Algorithms incorporating relation-aware aggregation, such as KGCN and \M, achieve the best performance in their respective settings, confirming the effectiveness of introducing relational attention mechanisms.
\end{itemize}

Overall, our framework outperforms existing federated learning algorithms and achieves competitive performance compared to centralized algorithms.

\subsection{Sensitivity Analysis}
\subsubsection{Activated client number}
In general, a smaller number of activated clients in each training round will speed up the model convergence and conversely better capture the global user data distribution. We test the algorithm on three datasets with three different numbers of activation clients, and the results are shown in the figure above. Probably due to the sparse data and a large number of clients, a small adjustment has a limited impact on the final results and the Last.FM and Book-Crossing datasets both show a small decrease in AUC when 64 clients are activated.
\subsubsection{Receptive field depth}
By testing different receptive field depths, we note that an excessive receptive field reduces model prediction accuracy. As data sparsity decreases, better performance needs a larger receptive field, while a one-layer perceptual region is sufficient to achieve better performance on those sparse data sets.

\begin{figure}[ht]
  \centering
  \subfigure{
    \includegraphics[width=0.35\textwidth]{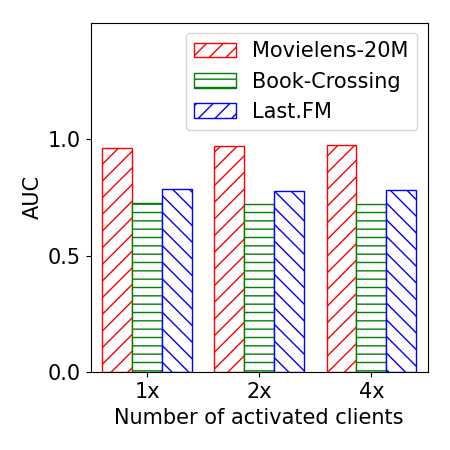}
    \label{fig:subfig2}
  }
  \subfigure{
    \includegraphics[width=0.35\textwidth]{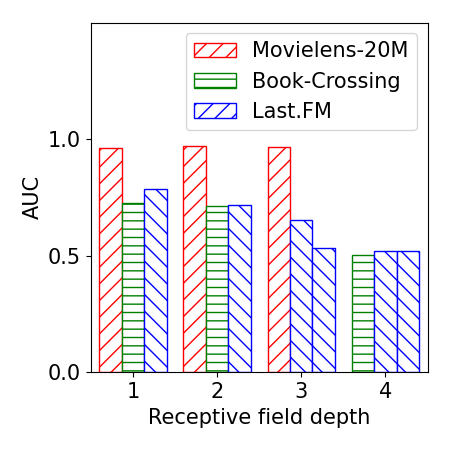}
    \label{fig:subfig2}
  }\\
  \vspace{-1.0em}
  \caption{Sensitivity analysis of activated clients and receptive field depth.}
  \label{fig:hyperparameter}
\end{figure}

\subsubsection{Interaction item protection}
\begin{figure}[t]
  \centering
  \includegraphics[width=\linewidth]{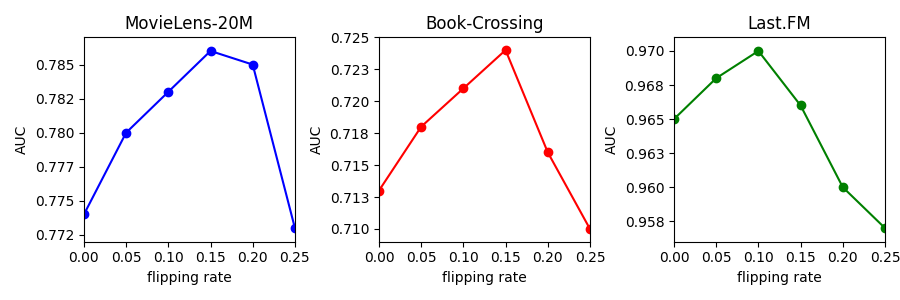}
  
  \vspace{-1.0em}
  \caption{Effect of flipping rate on AUC.}
  \label{fig:flip}
  \vspace{-1.5em}
\end{figure}

We introduce new interaction record protection and assess diverse flipping rates, with corresponding results depicted in Fig.~\ref{fig:flip}. 
Generally, integrating privacy-preserving mechanisms often diminishes recommendation accuracy.
However given limited client-side graph data, our scenario tends to induce model overfitting. 
Hence, proper regularization effectively enhances recommendation accuracy and privacy protection. Notably, excessive flip rates can compromise system performance despite heightened privacy. Our experiments indicate a balance between accuracy and privacy at a flipping rate of 0.1.

\section{Conclusion}
This paper introduces a novel federated learning framework, \M, which employs GNN for recommendation tasks. Our approach integrates KG information while upholding user privacy. The limitation here is the absence of user connections, and our forthcoming focus is on improving the efficiency and interpretability of utilizing existing user connections without introducing new private data.
Specifically, a server-side KG is created from public item data, maintaining relevant embeddings. The client conceals local interaction items and requests server training data. The server samples a KG subgraph and distributes it with the GNN model to the client. The client then expands its user-item graph with the KG subgraph for training, uploading the gradient for server aggregation. Our framework creates higher-order interactions without extra privacy data, relying solely on public information for KG. Sampled KG subgraphs enhance local training by capturing interactions between users and items without direct links. We employ LDP and pseudo-labeling to protect privacy and reduce overhead by requesting partial data. Gradients are encrypted using LDP for user preference protection and local user embedding storage.
Experimental results on three datasets demonstrate our framework's superiority over SOTA federated learning recommendation methods. It also performs competitively against centralized algorithms while preserving privacy.

\subsubsection{Acknowledgements} This work is supported by the National Key Research and Development Program of China under Grant No.2021YFB1714600 and the National Natural Science Foundation of China under Grant No.62072204 and No.62032008. The computation is completed in the HPC Platform of Huazhong University of Science and Technology and supported by the National Supercomputing Center in Zhengzhou.

\bibliographystyle{unsrt}
\bibliography{neurips_2023}

\begin{thebibliography}{10}

\bibitem{wang2019kgat}
Xiang Wang, Xiangnan He, Yixin Cao, Meng Liu, and Tat-Seng Chua.
\newblock Kgat: Knowledge graph attention network for recommendation.
\newblock In {\em Proceedings of the 25th {ACM} {SIGKDD} International Conference on Knowledge Discovery and Data Mining, {KDD}}, pages 950--958, 2019.

\bibitem{guo2020survey}
Qingyu Guo, Fuzhen Zhuang, Chuan Qin, Hengshu Zhu, Xing Xie, Hui Xiong, and Qing He.
\newblock A survey on knowledge graph-based recommender systems.
\newblock {\em IEEE Transactions on Knowledge and Data Engineering}, 34(8):3549--3568, 2020.

\bibitem{zhou2020graph}
Jie Zhou, Ganqu Cui, Shengding Hu, Zhengyan Zhang, Cheng Yang, Zhiyuan Liu, Lifeng Wang, Changcheng Li, and Maosong Sun.
\newblock Graph neural networks: A review of methods and applications.
\newblock {\em AI Open}, 1:57--81, 2020.

\bibitem{wu2022graph}
Shiwen Wu, Fei Sun, Wentao Zhang, Xu~Xie, and Bin Cui.
\newblock Graph neural networks in recommender systems: a survey.
\newblock {\em ACM Computing Surveys}, 55(5):1--37, 2022.

\bibitem{voigt2017eu}
Paul Voigt and Axel Von~dem Bussche.
\newblock The {EU} general data protection regulation ({GDPR}).
\newblock {\em A Practical Guide, 1st Ed., Cham: Springer International Publishing}, 10(3152676):10--5555, 2017.

\bibitem{chai2020secure}
Di~Chai, Leye Wang, Kai Chen, and Qiang Yang.
\newblock Secure federated matrix factorization.
\newblock {\em IEEE Intelligent Systems}, 36(5):11--20, 2020.

\bibitem{wu2021fedgnn}
Chuhan Wu, Fangzhao Wu, Yang Cao, Yongfeng Huang, and Xing Xie.
\newblock {FedGNN}: Federated graph neural network for privacy-preserving recommendation.
\newblock {\em arXiv preprint arXiv:2102.04925}, 2021.

\bibitem{liu2022federated}
Zhiwei Liu, Liangwei Yang, Ziwei Fan, Hao Peng, and Philip~S Yu.
\newblock Federated social recommendation with graph neural network.
\newblock {\em ACM Transactions on Intelligent Systems and Technology}, 13(4):1--24, 2022.

\bibitem{wang2020ckan}
Ze~Wang, Guangyan Lin, Huobin Tan, Qinghong Chen, and Xiyang Liu.
\newblock Ckan: collaborative knowledge-aware attentive network for recommender systems.
\newblock In {\em Proceedings of the 43rd International {ACM} {SIGIR} Conference on Research and Development in Information Retrieval, {SIGIR}}, pages 219--228, 2020.

\bibitem{sun2020multi}
Rui Sun, Xuezhi Cao, Yan Zhao, Junchen Wan, Kun Zhou, Fuzheng Zhang, Zhongyuan Wang, and Kai Zheng.
\newblock Multi-modal knowledge graphs for recommender systems.
\newblock In {\em Proceedings of the 29th ACM International Conference on Information \& Knowledge Managemen, {CIKM}}, pages 1405--1414, 2020.

\bibitem{wang2019knowledge}
Hongwei Wang, Miao Zhao, Xing Xie, Wenjie Li, and Minyi Guo.
\newblock Knowledge graph convolutional networks for recommender systems.
\newblock In {\em Proceedings of the World Wide Web Conference, {WWW}}, pages 3307--3313, 2019.

\bibitem{wang2019knowledge2}
Hongwei Wang, Fuzheng Zhang, Mengdi Zhang, Jure Leskovec, Miao Zhao, Wenjie Li, and Zhongyuan Wang.
\newblock Knowledge-aware graph neural networks with label smoothness regularization for recommender systems.
\newblock In {\em Proceedings of the 25th {ACM} {SIGKDD} International Conference on Knowledge Discovery and Data Mining, {KDD}}, pages 968--977, 2019.

\bibitem{JinBYDGYS23}
Hai Jin, Dongshan Bai, Dezhong Yao, Yutong Dai, Lin Gu, Chen Yu, and Lichao Sun.
\newblock Personalized edge intelligence via federated self-knowledge distillation.
\newblock {\em {IEEE} Transactions on Parallel and Distributed Systems}, 34(2):567--580, 2023.

\bibitem{zhang2021subgraph}
Ke~Zhang, Carl Yang, Xiaoxiao Li, Lichao Sun, and Siu~Ming Yiu.
\newblock Subgraph federated learning with missing neighbor generation.
\newblock In {\em Proceedings of the Annual Conference on Neural Information Processing Systems, {NeurIPS}}, volume~34, pages 6671--6682, 2021.

\bibitem{peng2021differentially}
Hao Peng, Haoran Li, Yangqiu Song, Vincent Zheng, and Jianxin Li.
\newblock Differentially private federated knowledge graphs embedding.
\newblock In {\em Proceedings of the 30th ACM International Conference on Information \& Knowledge Management, {CIKM}}, pages 1416--1425, 2021.

\bibitem{ammad2019federated}
Muhammad Ammad-Ud-Din, Elena Ivannikova, Suleiman~A Khan, Were Oyomno, Qiang Fu, Kuan~Eeik Tan, and Adrian Flanagan.
\newblock Federated collaborative filtering for privacy-preserving personalized recommendation system.
\newblock {\em arXiv preprint arXiv:1901.09888}, 2019.

\bibitem{10109123}
Guoren Wang, Yue Zeng, Rong-Hua Li, Hongchao Qin, Xuanhua Shi, Yubin Xia, Xuequn Shang, and Liang Hong.
\newblock Temporal graph cube.
\newblock {\em IEEE Transactions on Knowledge and Data Engineering}, pages 1--15, 2023.

\bibitem{cao2022geometric}
Wenming Cao, Canta Zheng, Zhiyue Yan, and Weixin Xie.
\newblock Geometric deep learning: progress, applications and challenges.
\newblock {\em Science China Information Sciences}, 65(2):126101, 2022.

\bibitem{tian2022fedbert}
Yuanyishu Tian, Yao Wan, Lingjuan Lyu, Dezhong Yao, Hai Jin, and Lichao Sun.
\newblock {FedBERT}: When federated learning meets pre-training.
\newblock {\em ACM Transactions on Intelligent Systems and Technology}, 13(4):1--26, 2022.

\bibitem{harper2015movielens}
F~Maxwell Harper and Joseph~A Konstan.
\newblock The movielens datasets: History and context.
\newblock {\em {ACM} Transactions on Interactive Intelligent Systems}, 5(4):19:1--19:19, 2016.

\bibitem{ziegler2005improving}
Cai-Nicolas Ziegler, Sean~M McNee, Joseph~A Konstan, and Georg Lausen.
\newblock Improving recommendation lists through topic diversification.
\newblock In {\em Proceedings of the World Wide Web Conference, {WWW}}, pages 22--32, 2005.

\bibitem{cantador2011second}
Iv{\'a}n Cantador, Peter Brusilovsky, and Tsvi Kuflik.
\newblock Second workshop on information heterogeneity and fusion in recommender systems ({HetRec2011}).
\newblock In {\em Proceedings of the 2011 {ACM} Conference on Recommender Systems, {RecSys}}, pages 387--388, 2011.

\bibitem{koren2008factorization}
Yehuda Koren.
\newblock Factorization meets the neighborhood: a multifaceted collaborative filtering model.
\newblock In {\em Proceedings of the 14th {ACM} {SIGKDD} International Conference on Knowledge Discovery and Data Mining, {KDD}}, pages 426--434, 2008.

\bibitem{rendle2012factorization}
Steffen Rendle.
\newblock Factorization machines with libfm.
\newblock {\em {ACM} Transactions on Intelligent Systems and Technology}, 3(3):1--22, 2012.

\bibitem{yu2014personalized}
Xiao Yu, Xiang Ren, Yizhou Sun, Quanquan Gu, Bradley Sturt, Urvashi Khandelwal, Brandon Norick, and Jiawei Han.
\newblock Personalized entity recommendation: A heterogeneous information network approach.
\newblock In {\em Proceedings of the 7th {ACM} International Conference on Web Search and Data Mining, {WSDM}}, pages 283--292, 2014.

\bibitem{zhang2016collaborative}
Fuzheng Zhang, Nicholas~Jing Yuan, Defu Lian, Xing Xie, and Wei-Ying Ma.
\newblock Collaborative knowledge base embedding for recommender systems.
\newblock In {\em Proceedings of the 22nd {ACM} {SIGKDD} International Conference on Knowledge Discovery and Data Mining, {KDD}}, pages 353--362, 2016.

\bibitem{wang2018ripplenet}
Hongwei Wang, Fuzheng Zhang, Jialin Wang, Miao Zhao, Wenjie Li, Xing Xie, and Minyi Guo.
\newblock Ripplenet: Propagating user preferences on the knowledge graph for recommender systems.
\newblock In {\em Proceedings of the 27th {ACM} International Conference on Information and Knowledge Management, {CIKM}}, pages 417--426, 2018.

\end{thebibliography}


\end{document}